\documentclass[12pt]{article}
\usepackage{amssymb,amsmath,slashed}
\usepackage[pdftex]{graphicx}

\makeatletter

\@addtoreset{equation}{section}
\def\section{\@startsection {section}{1}{\z@}{-2.5ex plus -1ex minus
 -.2ex}{1.3ex plus .2ex}{\large\bf}}
\def\subsection{\@startsection{subsection}{2}{\z@}{-2.25ex plus%
 -1ex minus -.2ex}{0.5ex plus .2ex}{\bf}}

\advance \voffset by -0.8in
\advance \hoffset by -0.6in
\def\Dslash{\slashed{D}}
\def\bone{{\mbox{\bf 1}}_2}

\def\lf{\sigma}

\def\Ric{\mbox{Ric}}

\newcommand{\CP}{\mathbb{CP}}

\def\bpm{\begin{pmatrix}}
\def\epm{\end{pmatrix}}

\newcommand{\RR}{\mathbb{R}}
\newcommand{\EE}{\mathbb{E}}
\newcommand{\CC}{\mathbb{C}}

\def\bee{\begin{equation}}
\def\eee{\end{equation}}

\begin{document}
\begin{flushright}
EMPG-14-21
\end{flushright}
\begin{center}
{\Large \bf  Time evolution in a geometric model of a particle }

\baselineskip 18pt

\vspace{0.4 cm}

M.~F.~Atiyah, \\
Maxwell Institute for Mathematical Sciences and School of Mathematics, \\
University of Edinburgh,
Kings Buildings,  Edinburgh EH9 3JZ, UK \\
{\tt M.Atiyah@ed.ac.uk} 
\vspace{0.4cm}

 G.~Franchetti and B.~J.~Schroers,\\
Maxwell Institute for Mathematical Sciences and Department of Mathematics, \\
Heriot-Watt University,
Riccarton, Edinburgh EH14 4AS, UK  \\
 {\tt g.franchetti@hw.ac.uk} and  {\tt b.j.schroers@hw.ac.uk} \\
\vspace{0.4cm}

{18 December  2014} 
\end{center}

\begin{abstract}
\noindent We analyse the properties of a (4+1)-dimensional Ricci-flat spacetime which may be viewed as an evolving Taub-NUT geometry,  and give exact solutions of the Maxwell and gauged  Dirac equation on this background. We interpret these solutions in terms of  a  geometric model of the electron and its spin, and discuss links  between the resulting picture and Dirac's Large Number Hypothesis.
\end{abstract}

\baselineskip 16pt
\parskip 2.5 pt
\parindent 8pt


\section{Introduction}

This paper  continues the exploration of a purely geometrical description of elementary particles begun in  \cite{AMS} by including  time. We  focus on one particular model, review its geometry in detail and exhibit a time-dependent version. Our review includes new insights  into the description of spin in the geometric framework which build  on  results obtained in \cite{JS}. 

The work reported here was inspired by  Dirac's two-metric formalism, introduced in the context of his Large Number Hypothesis (LNH) \cite{Dirac-1, Dirac0, Dirac1,Dirac2}. The two metrics considered by Dirac --- the usual spacetime metric in Einstein's theory of gravitation and a second  one for the description of `atomic' phenomena --- are related in a way which is echoed in the relation between the spatial and fibre geometry of   the Taub-NUT space. Extending this relation to a spacetime picture led us to the   (4+1)-dimensional Ricci-flat extension of  the Taub-NUT space which we shall discuss, and which had previously been derived in \cite{GLP}.

In the geometric approach  of  \cite{AMS},  static particles are modelled by four-manifolds, and conserved quantum numbers of particle physics are interpreted in terms of  topological invariants of four-manifolds. The models for  electrically charged particles are non-compact four-manifolds  which  are asymptotically fibred by circles and, for  asymptotic  $U(1)$ fibrations, the Chern number  is identified with  minus the electric charge. The asymptotic picture is essentially the electric-magnetic dual of the Kaluza-Klein description of electromagnetism. 

 In  \cite{AMS},  Euclidean Taub-NUT   space  (TN) is proposed as a geometric model of the electron.  TN is more conventionally interpreted as the  Kaluza-Klein geometrisation of a magnetic monopole \cite{Sorkin,GP}, and the results of our paper are interesting when viewed from this perspective, too. However, for definiteness we adopt the dual interpretation given  in  \cite{AMS} in the following discussion. 
 
Up to scale,  TN is  the unique complete  hyperk\"ahler four-manifold  with  isometry group $U(2)$. With the complex structure defined by the central $U(1)$ subgroup of $U(2)$, it is naturally isomorphic to $\CC^2$. The hyperk\"ahler structure induces a  map  from TN to Euclidean three-space $\EE^3$ through  its  moment map.  The moment map  intertwines the  $SU(2)$ action on  TN with the $SO(3)$ action on $\EE^3$.

The central $U(1)$ action has one fixed point, called the NUT. Away from the NUT,  the 
TN  geometry has  the structure of a $U(1)$ bundle, and 
the hyperk\"ahler moment map is the projection of this fibration. In the interpretation of TN as a model for the electron,  the  Chern number of the fibration is  the negative of the electric charge and the  radius of the  fibres over  infinity in $\EE^3$ is the classical electron radius. The  model  captures the point-like nature of the electron  through the unique fixed point of the $U(1)$ action, but does so in the context of  a smooth geometry with an associated length scale.

 In \cite{AMS}, it was proposed that the spin degrees of elementary particles may be captured in geometric models through the zero-modes of the Dirac operator on the relevant four-manifold. This was elaborated in \cite{JS} where the zero-modes of the Dirac operator on TN minimally coupled to an abelian connection were studied in detail, building on earlier  papers by Pope \cite{Pope1,Pope2}. Up to an overall factor $p$, the curvature of the abelian connection is the unique harmonic and square integrable two-form on TN. 
 The space of zero modes turns out to combine all irreducible $SU(2)$ representations  up to a dimension determined by $p$ \cite{JS}.

Here we begin,  in Sect.~\ref{spinsect}, by revisiting the TN model for the electron, including its spin. The scale parameter in the TN geometry  plays a crucial role in our discussion and   we explain its relation to the asymptotic size of the $U(1)$ fibres. We emphasise two distinct geometrical aspects of TN space, namely its $U(1)$ fibres,   which we interpret as microscopic  and having a radius comparable  to the classical electron radius,    and its three-dimensional base which we interpret as  macroscopic  and referring to positions in ordinary Euclidean space. 

Using the explicit forms of the zero-modes  given in  \cite{JS},  we then exhibit a  natural link between the  spin 1/2 zero-modes of the Dirac operator and complex coordinates on TN  defined in terms of the  $U(1)$ fibre. The resulting picture is one where  TN is viewed microscopically as spin space  and macroscopically as   a $U(1)$-bundle over the  position space $\EE^3$ of a non-relativistic electron.

 Our time-dependent model is introduced and discussed in  Sect.~\ref{timesect}. It involves a remarkably simple solution of the (4+1)-dimensional Einstein equations  first given in \cite{GLP} which can be interpreted as an evolving TN geometry with the scale parameter  changing in  time. Surprisingly,  allowing the   harmonic two-form and the zero-modes of the Dirac operator on TN to change adiabatically with time gives {\em exact} solutions of the Maxwell and Dirac equations on the (4+1)-dimensional background provided the coefficient  $p$ of the abelian connection  is kept constant at an integer value. Thus, our interpretation of zero-modes as spin 1/2 states of the electron carries over to the time-dependent setting. 
  
  During the time evolution, the microscopic length scale decreases relative  to the macroscopic length scale of the TN geometry. As anticipated,  our model thereby becomes  a natural illustration of Dirac's  Large Number Hypothesis \cite{Dirac-1, Dirac0, Dirac1,Dirac2}. This is explained and elaborated in our final  Sect.~\ref{concl} where we discuss the interpretation of the evolving TN geometry as a model for the electron as well as  possible generalisations.

\section{Spin and position in the Taub-NUT geometry}
\label{spinsect}

\subsection{Notation and conventions}

For a more detailed discussion  and to fix our conventions we describe the TN geometry  in terms of $SU(2)$-orbits and a transverse coordinate $r$. We parametrise $h\in SU(2)$ further  in terms of a pair of complex numbers $z=(z_1,z_2)$  as 
\bee
h= \begin{pmatrix}
z_1  & -\bar{z}_2\\
z_2 &
\phantom{-} \bar{z}_1 \end{pmatrix}, \quad  |z_1|^2 + |z_2|^2= 1,
\eee
and introduce  $su(2)$ generators   $t_j=-\frac i2 \tau_j$, where $\tau_j$, $j=1,2,3$, are  the  Pauli matrices.
Then defining
left-invariant one-forms $\sigma_j$ via
\bee
h^{-1}dh = t_1\sigma_1+t_2\sigma_2+t_3\sigma_3, 
\eee
the TN metric is  of the general  Bianchi IX form
\bee
\label{athi}
 ds^2 = f^2dr^2 + a^2\lf_1^2 +b^2\lf_2^2 +c^2\lf_3^2. 
\eee

In four dimensions, the hyperk\"ahler property is equivalent to the self-duality of the Riemann tensor with respect to an orientation which is opposite to that defined by the hyperk\"ahler complex structures. As explained in detail in \cite{AMS}, this is  in fact the complex structure defined by the central $U(1)$ subgroup of the isometry group $U(2)$.  In our conventions for the left-invariant one-forms on $SU(2)$   (which are those of \cite{JS}  but have the opposite sign of the forms used in \cite{AMS}, so satisfy $d\sigma_3=\sigma_2\wedge \sigma_1$) the orientation is given by the following co-frame 
\bee
\label{tetrad}
e_1= a \lf_1, \quad e_2= b \lf_2, \quad e_3=c\lf_3, \quad e_4 =-f dr,
\eee
and associated volume element
\begin{align}
\label{volume}
 e_1\wedge e_2\wedge e_3\wedge e_4  =fabc  \, dr \wedge \lf_1 \wedge \lf_2\wedge \lf_3.
 \end{align}

For a self-dual metric, the coefficient functions satisfy the self-duality equations 
\bee
\label{dual}
 \frac {2bc} {f}\frac { da}{dr} = (b-c)^2 -a^2,\; \; \text{+ cycl.},
\eee
where `+ cycl.' means we add the two further equations
obtained by cyclic permutation of $a,b,c$.   For the TN solution, the central $U(1)$ symmetry imposes $a=b$, so the equation for $c$ becomes
\bee
\label{crform}
 \frac { dc}{dr} =  -\frac{fc^2}{2a^2}. 
\eee

We can fix the coordinate $r$ transverse to the $SU(2)$ orbits by picking any nowhere vanishing function $f$. There are two  geometrically natural choices which are important for us. 

\subsection{Microscopic or spin coordinates}
One natural choice  is to pick $f= -2a^2/c^2$, so that $r=c$. 
Geometrically, $2c$ is the radius of the $U(1)$ fibres, the factor of two stemming from the  angular range $[0,4\pi)$ of a $U(1)$ subgroup of $SU(2)$. To write the metric in terms of  this coordinate, we 
 divide  the equation for $a$ by the equation for  $c$ in \eqref{dual}   to find
 \bee
 \label{homoeq}
\frac{da}{dc} =-  \frac a c   + 2\left(\frac{a}{c}\right)^2.
\eee

The TN solution has $a=0$ when $c=0$, but this does not fix the solution of  \eqref{homoeq} since $c=0$ is a singular point. Instead, 
the general solution is 
\bee
\label{acform}
a= \frac{ c}{1-\frac{c^2}{\Lambda^2}},
\eee
and involves an arbitrary positive constant  $\Lambda$  which fixes the range of  $c$ as $ [0,\Lambda)$
 and the asymptotic radius  of the $U(1)$ fibres as  $2\Lambda$. There is, therefore, a one-parameter family of TN metrics, labelled by the size of the asymptotic circle. The  metric  depends on $\Lambda$ and takes the form
\bee
\label{TNc}
ds^2= \frac{4}{(1-\frac{c^2}{\Lambda^2})^4} dc^2 + \frac{c^2 }{(1  -\frac{c^2}{\Lambda^2})^2} (\lf_1^2 +\lf_2^2) + c^2 \lf_3^2.
\eee

There is  a family of `cigar shaped' geodesic submanifolds which are invariant under  the global $U(1)$ action,  parametrised by the  coset $SU(2)/U(1)$ which, as we shall see,  is the  two-sphere of spatial directions in the base  $\EE^3$. The metric on each submanifold is 
\bee
ds^2 = \frac{4}{(1-\frac{c^2}{\Lambda^2})^4} dc^2 + c^2 d\gamma^2,
\eee
where $\gamma$ is an angular coordinate  with range $[0,4\pi)$. The Gauss curvature at the tip of the cigar (the NUT in TN) is 
\bee
\label{Gauss}
 K = \frac{1}{\Lambda^2}.
 \eee
The curvature radius at the  NUT is  half the asymptotic radius of the fibres, showing  that the asymptotic scale $\Lambda$ can also be read off from the geometry near the NUT.

For later use we note that, with $c$ as coordinate, the proper radial distance is 
\bee
\rho(c) =  \int_0^c \frac{2 ds}  { \left(1-\frac{s^2}{\Lambda^2}\right)^2}  = 
 \frac{c}{ \left(1-\frac{c^2}{\Lambda^2}\right) }+ \Lambda \tanh^{-1}\left(\frac{c}{\Lambda}\right), 
 \eee
and that this relation is linear $\rho(c)\approx 2c$ for small $c$.

We can use the fibre radius $2c$ to define global complex coordinates on TN via 
 \bee 
 \label{wcoord}
 w=2cz.
 \eee 
 Then $w=(w_1,w_2)$  takes values in the  open ball $B_\Lambda  =\{w  \in \CC^2| \;|w|< 2\Lambda\}  $ of radius $2\Lambda$. We refer to $w$ as a microscopic coordinate   since its  magnitude  is related to the fibre radius whose  asymptotic value  $\Lambda$ was interpreted as the classical electron radius in \cite{AMS}. Since  $\frac 1 4 ( \lf_1^2+\lf_2^2 +\lf_3^2)$ is the round metric on the unit three-sphere, the TN metric near the NUT is   flat:
 \bee
 \label{atomic}
 ds^2\approx  |dw_1|^2 +|dw_2|^2. 
 \eee

As we shall explain in Sect.~\ref{spinrev}, the coordinates $w_1$ and $w_2$ are  naturally interpreted as spin coordinates  for TN. They obviously transform in the fundamental representation of $SU(2)$, but more specifically we shall show that the spin 1/2 zero-modes of a twisted Dirac operator on TN  are  linear functions   of them.

\subsection{Macroscopic or position coordinates} 
A second, geometrically natural choice of radial coordinate is provided by the $U(1)$ fibration of TN  over $\EE^3$ away from the origin.  For any fixed $SU(2)$ orbit, this is 
the Hopf fibration 
\bee
\pi:S^3\rightarrow S^2, \quad z\mapsto \vec{n} =z^\dagger \vec{ \tau} z.
\eee
Using the  Euler angle parametrisation of $SU(2)$ given in \cite{JS}, 
\bee
\label{zdeff}
z_1= e^{-\frac i 2 (\alpha+ \gamma)} 
\cos \frac{\beta}{2} , \quad 
z_2=
e^{\frac i 2 (\alpha - \gamma)} \sin \frac{\beta }{2}, \quad \beta \in[0,\pi), \alpha\in [0,2\pi), \gamma \in [0,4\pi), 
\eee
we have $
\vec{ n}= (\sin\beta\cos\alpha, \sin\beta \sin \alpha,\cos\beta)  $
and  so
\bee
\lf_1^ 2+\lf_2^2  = d\beta^2 + \sin^2\beta d\alpha^2
\eee
is the standard round metric on the unit two-sphere parametrised by $\vec{n}$. The choice
\bee
f=-\frac{a}{r},
\eee
then makes the metric  isotropic   in the coordinate vector 
\bee
\vec{x}=r \vec{n}, 
\eee
i.e.,  it brings it into the form
\bee
\label{TNGH}
ds^2 = \frac{a^2}{r^2} d\vec{x}^2 + c^2 \lf_3^2.
\eee

Observe that $r$, like  $a$ and $c$,  necessarily  has  the dimension  length. To determine $a$ as a function of $r$ we substitute  \eqref{acform} into \eqref{crform}  and integrate:
\bee
\label{crrelation}
\frac{d}{dr} \left(\frac{r}{c^2} \right)=\frac{1}{\Lambda^2} \; \Leftrightarrow \; \frac{1}{c^2} =\frac{1}{\Lambda^2} + \frac{1}{rL}.
\eee
The solution requires an integration constant of dimension inverse length which we  have denoted by   $1/L$. Thus,  defining the dimensionless quantitity
\bee
\label{epdef}
\epsilon = \frac{L^2}{\Lambda^2},
\eee
and introducing 
\bee
\label{potential}
V=\epsilon+\frac L r,
\eee
we have, from \eqref{acform},
\bee
\label{GH}
c = \frac{L}{\sqrt{V}}, \qquad a = r\sqrt{V},
\eee
with the radial coordinate $r$  taking values in $[0,\infty)$.  

The metric \eqref{TNGH} with the expressions \eqref{GH} for $a$ and $c$  is the TN metric
\bee
\label{TNexplicit}
ds^2= \left(\epsilon +\frac L r\right) d\vec{x}^2 + \frac {rL^2}{\epsilon r +L} \sigma_3^2
\eee
  in Gibbons-Hawking form. We have re-derived it here to clarify the origin and the interpretation of the two constants $\epsilon$ and $L$ which appear in it. The length scale $\Lambda>0$ fixes   the asymptotic  radius of the $U(1)$ fibres. The second length scale $L$ appears when we introduce coordinates on the base $\EE^3$ of the fibration. For the regular (`positive mass') TN geometry one needs $L>0$ and we assume this in the following.
The dimensionless quantity $\sqrt{\epsilon}$ is the ratio  of these length scales. Only the scale $\Lambda$ of the fibre has an invariant meaning. This can also be seen by expressing the  TN metric  \eqref{TNexplicit} in terms of the   TN metric with $\epsilon=L=1$  via a  change of coordinates  $\tilde{r}= (\epsilon r)/L$ on the base and an overall re-scaling by $\Lambda^2$:
\bee
ds^2=  \frac{L^2}{\epsilon}\left[ \left(1+\frac{1}{\tilde{r}}\right)d\tilde{\vec{x}}^2 + \frac{\tilde{r}}{\tilde{r}+1 }\sigma_3^2\right].
\eee

Defining $R/2=\sqrt{Lr}$ as the geometric mean of $r$ and the scale parameter $L$, so that 
\bee
\label{rRdef}
r=\frac {1} {4L} R^2, 
\eee
we introduce macroscopic complex coordinates on TN via
\bee
\label{bigw}
W=Rz \in \CC^2. 
\eee
They project to $\vec{x}$ via the Hopf projection 
\bee
\vec{x} = \frac{1}{4L}W^\dagger \vec{\tau} W. 
\eee
Observe that for small $r$,  from  \eqref{crrelation}  and \eqref{rRdef}, 
 \bee
 \label{rcrel}
 r\approx \frac {1} {L} c^2, \qquad R\approx 2c,  \qquad w\approx W,
\eee
so that  microscopic and macroscopic coordinates  agree near the NUT.

As advertised in the Introduction, 
the coordinate vector $\vec{x}$ for the base space can be obtained more invariantly from the hyperk\"ahler structures of TN. The three hyperk\"ahler forms $\omega_i$, $i=1,2,3$, are invariant under the $U(1)$ action. Writing 
\bee
X=\frac{\partial}{\partial \gamma}
\eee
for the vector field generating the $U(1)$ action, we have ${\mathcal L}_X \omega_i=0$, so that the forms $\iota_X\omega_i$ are closed.  
For the TN metric in the form \eqref{TNexplicit}, the hyperk\"ahler forms are 
\bee
\omega_i = L dx_i\wedge \lf_3 +\frac 1 2 V\epsilon_{ijk} d x_j\wedge d x_k, \qquad i=1,2,3.
\eee
Note that the orientation associated with the hyperk\"ahler structures
\bee
\omega_1\wedge\omega_1= \omega_2\wedge\omega_2=\omega_3\wedge\omega_3=
L V dx_1\wedge dx_2\wedge dx_3\wedge \sigma_3
\eee
is the opposite of our orientation \eqref{volume} which is determined by the global $U(1)$ action, as expected.
 Defining  associated  moment maps $\mu_i$  via
\bee
\iota_X \omega_i= - d\mu_i,
\eee
we  find that they are, up to the scale $L$ and   additive constants $q_i$,  the Euclidean position coordinates:
\bee
\mu_i =Lx_i + q_i, \qquad i=1,2,3. 
\eee

\subsection{The twisted Dirac operator on TN space}
\label{spinrev}

Even though TN is topologically trivial, it has non-trivial $L^2$-cohomology in the middle dimension \cite{Pope1}. The harmonic representative, unique up to scale,  is  the  exterior derivative of  the  one-form $c^2 \lf_3$ (which is not square-integrable). Here we want to interpret the harmonic representative as the curvature of an abelian connection on TN. We  adapt the conventions of \cite{JS} (where $\epsilon=1$) and write the connection one-form as 
\bee
\label{connection}
A = \frac{i p}{2} \frac{c^2}{\Lambda^2} \lf_3= \frac{i p}{2} \frac{\epsilon r}{\epsilon r +L} \lf_3,
\eee
so that the curvature is 
\bee
\label{curvature}
F=dA = \frac{ i\epsilon p}{2} \left(\frac{L}{(\epsilon r +L)^2} dr \wedge \sigma_3 
+\frac{r}{\epsilon r +L} \sigma_2\wedge \sigma_1 \right).
\eee
The coefficient $p$ can take arbitrary real values since TN has no non-trivial two-cycles. However, TN can be compactified to $\CP^2$ by adding a $\CP^1$  representing `spatial infinity'  \cite{AMS}.  In our conventions, this is the $\CP^1$ with homogeneous coordinate $w$.  If one requires that   $A$ extends to a connection on $\CP^2$ then $p$ has to take integer values, but we will not assume this in the following.  
 
 The  index of the Dirac operator on TN minimally coupled to the connection $A$ was computed by Pope in  \cite{Pope1,Pope2}  as  $\frac 1 2 [|p|] ([|p|]+1)$,
where $[x]$ is the largest integer {\em strictly} smaller than  the positive real number $x$. This index and the properties of Dirac  zero-modes are  studied in more detail in  the paper \cite{JS}  the   notation of which  we use here.  The  Dirac operator has the form 
\bee
\label{TNdirac}
\Dslash_p = \bpm 0 & T_p^{\dagger} \cr T_p & 0 \epm,
\eee
and all its normalisable zero-modes come from normalisable zero-modes of $T_p$. As explained in \cite{JS},  the space of all zero-modes decomposes into the sum of irreducible representations of $SU(2)$ with spins $j$ satisfying $2j+1 < |p|$. 

Fixing $p>0$ for definiteness, the zero-modes depend holomorphically on the complex coordinates. For fixed $j$, they can be written as 
\bee
\label{holosol}
\Psi_j(r,z_1,z_2) = \begin{pmatrix}
R_j(r)\sum_{m=-j}^ja_m z_1^{j-m}z_2^{j+m} \\
0\\
0 \\
0
\end{pmatrix}, 
\eee
where  $a_m$, $ m=-j,-j+1,\ldots, j-1, j$,  are  arbitrary complex constants. The  radial dependence  is determined by the differential equation 
\bee
\label{jq}
\left(L\partial_r+\frac \epsilon  2 \left(  p-(2j+1) \right)+ \left ( \frac 1 2 -j \right)\frac L r   -\frac {L^2}{2r(\epsilon  r+L)}\right)R_j(r)=0, 
\eee
which  has the general solution
\bee
\label{posq}
R_j(r )=C\frac{r^j}{ \sqrt{\epsilon r +L} }e^{((2j+1)-p)\frac{\epsilon r}{2L} }.
\eee
Here $C$ is a normalisation constant and  $j$ is the spin,  required to satisfy $ (2j+1) <p$ for normalisability.

In \cite{JS}, the $j=1/2$ doublet of states  was proposed as a model for  the spin degrees in  the TN model.  It was also pointed out that,  with the choice $p=2$,  the spin 1/2 states are not square-integrable but  have the form of a vortex,  with   constant magnitude at spatial infinity.  Remarkably, the radial dependence is a multiple of the  fibre radius  $2c$  \eqref{GH} in the TN metric. As a result, the first component of the spinor in \eqref{holosol} is simply  a  linear function in the   global coordinate  $w$ \eqref{wcoord} on TN, so that we can write the zero-modes as 
\bee
\label{holosolw}
\Psi_{\frac 12 } (r,z_1,z_2) = \begin{pmatrix}
a_{-\frac 1 2} w_1 +  a_{\frac 1 2} w_2\\
0\\
0 \\
0
\end{pmatrix}.
\eee
This  justifies our earlier interpretation of $w_1$ and $w_2$   as  coordinates on spin space. 

\subsection{Scaling properties and the  Landau limit}

In the limit $\epsilon \rightarrow 0$, the $U(1)$ fibre decompactifies and TN space becomes flat $\RR^4$. The microscopic and macroscopic coordinates $w$  and $W$  now coincide since, when $\epsilon =0$, 
\bee
c=\frac{R}{2} =\sqrt{Lr}. 
\eee
The TN  metric is  flat and given by     $|dW_1|^2 + |dW_2|^2$ in this limit. For negative values of $\epsilon$, the metric coefficients \eqref{GH} still satisfy the self-duality equations but the metric  \eqref{TNexplicit} is  now minus the `negative mass' TN geometry,  which is singular at $r=-L/\epsilon$. This metric arises as the asymptotic form of the Atiyah-Hitchin metric on the monopole moduli space \cite{GM}. 

With our normalisation,  the  abelian connection \eqref{connection} and curvature \eqref{curvature} vanish in the limit $\epsilon \rightarrow 0$, but it is instructive to consider the limit 
\bee
\epsilon \rightarrow 0, \quad {\epsilon p}\rightarrow \tilde p \neq 0. 
\eee
The connection becomes
\bee
A=i\frac{\tilde pR^2}{8L^2} \sigma_3 = \frac{\tilde p}{8L^2}\left( W_1d\bar{W}_1+W_2d\bar{W}_2 -\bar{W}_1dW_1-\bar{W}_2dW_2\right).
\eee
so that the curvature is essentially  the K\"ahler form on $\CC^2$:
\bee
F= \frac{\tilde p}{4L^2}\left(d W_1 \wedge d\bar{W}_1+ d W_2\wedge \bar{W}_2\right).
\eee
Physically, this can be thought of as a constant magnetic field in both  the $W_1$ and  the $W_2$ plane. It is not surprising, therefore, that the Dirac zero-modes become  products of Landau ground states in each of the planes, with a  Gaussian radial function multiplying holomorphic polynomials in $W_1$ and $W_2$:
\bee
W_1^{j-m}W_2^{j+m} e^{- \tilde p\frac {|W|^2}{2 L}}.
\eee
This is the Landau limit of our model. All energy levels, including the zero-energy state,  have an infinite  degeneracy in this limit. 

\section{Introducing time}
\label{timesect}

Somewhat surprisingly, an affine time dependence of the scaling parameter $\epsilon$ gives rise to an exact solution of the (4+1)-dimensional vacuum Einstein equations. This  solution was first obtained  in \cite{GLP} as the Kaluza-Klein form of the electric-magnetic dual of  a  Kastor-Traschen solution of  the Einstein-Maxwell equations.   For our purposes, it  is instructive to derive it directly as follows. 

We write $g_{\text TN}(t)$ for the TN metric  \eqref{TNexplicit} 
with the parameter $\epsilon$ depending on an additional time variable $t$, i.e., we  consider the adiabatic variation of the potential
\bee
V=\epsilon(t) +\frac{L}{r}.
\eee
Then  the  metric 
\bee
\label{timeTN}
ds^2 = - dt^2+ g_{\text TN}(t)
\eee
has Ricci scalar
\bee
s = \frac {2r}{ (\epsilon(t) r+L)} \ddot{\epsilon}, 
\eee
with dots denoting derivatives with respect  to $t$. 
Moreover, in terms of the coefficient function \eqref{GH} and  the co-frame 
\bee
\label{ttetrad}
e^0= dt, \quad e^1= a \lf_1, \quad e^2= a \lf_2, \quad e^3=c\lf_3, \quad e^4 =\frac a r dr,
\eee
the  Ricci tensor has non-vanishing components
\bee
\Ric_{00}  = -\frac 1 2 s, \quad  \Ric_{11}= \Ric_{22}=  -\Ric_{33}= \Ric_{44}= \frac 1 4 s. 
\eee
Thus
\bee
\label{riccond}
\Ric=0 \Leftrightarrow \ddot{\epsilon} = 0,
\eee
so that  an affine dependence of $\epsilon$ on $t$ yields a Ricci-flat metric.
Obtaining an exact solution  of a time-dependent problem from an adiabatic ansatz  is rather unusual and it is not clear to us why it works in this instance. 

Concentrating on the simplest case 
\bee
\epsilon(t)=t,
\eee
 the family of metrics $g_{\text TN}(t)$ starts, at $t=0$, with Euclidean four-space. For $t>0$ one dimension compactifies to the  circle fibre  of TN space, with the radius of the circle  decreasing with $t$ as
\bee
\label{arrow}
\Lambda(t) =\frac{L}{\sqrt{t}}.
\eee
For $t<0$ the metric $g_{\text TN}(t)$ is the negative of the  (singular) negative mass TN metric, so that \eqref{timeTN} is a five-dimensional singular  space  whose signature flips from $(-,+,+,+,+)$ to $(-,-,-,-,-)$, with the two regimes separated by the singular region  $rt=-L$.  

In order to extend our   description of spin to the time-dependent case we look for   solutions of the Maxwell and Dirac equations on the (4+1)-dimensional space-time \eqref{timeTN}. We continue to work with the co-frame \eqref{tetrad}, use Greek  indices $\mu,\nu,\ldots$ in the range $0,\ldots, 4$ and raise or lower them with the `mostly plus' metric diag$(-1,1,1,1,1)$.  It is then straightforward to check that allowing $\epsilon$ and $p$  to vary with time,  the one-form 
\bee
\label{tconnection}
A = \frac{i p (t) }{2} \frac{\epsilon (t) r}{\epsilon (t) r +L} \lf_3
\eee
gives rise to the curvature 
\bee
F = dA =  \frac{i }{2}\left(\frac{(\dot \epsilon p - \dot p \epsilon)rL -   \dot p L^2 }{(\epsilon r +L)^2}
dt \wedge \sigma_3
+\frac{p\epsilon L}{(\epsilon r +L)^2} dr \wedge \sigma_3 
+\frac{p\epsilon r }{\epsilon r +L} \sigma_2\wedge \sigma_1\right), 
\eee
which  satisfies 
\bee
d\star F= 0,
\eee
provided that
\bee 
\ddot{\epsilon} =0 \quad \text{and} \quad \ddot{p}=0.
\eee
 In other words, evolving the abelian gauge field $A$ adiabatically with the TN metric gives a solution of the Maxwell equations. However, the energy-momentum tensor does not vanish, so we do not obtain a solution of the coupled Einstein-Maxwell equations. 

For the Dirac equation in 4+1 dimensions we use the gamma matrices
 \bee
\label{gamma4}
\gamma_0=\bpm \bone  &  0 \cr  0   & -\bone \epm, \qquad 
 \gamma_i =
\bpm 0 & \tau_j\cr -\tau_j& 0 \epm, \; \quad \gamma_4= \bpm 0 & -i\bone \cr -i\bone & 0 \epm .
\eee
The spin connection one-form for  \eqref{timeTN},
defined via
\bee
 de^\mu +\omega^{\mu}_{\;\;\nu} \wedge e^\nu =0,
\eee
has the four-dimensional Euclidean components given, for example, in \cite{JS} and the additional components
\begin{align}
\omega_{01} =-\frac{\dot \epsilon  r}{2\sqrt{\epsilon +\frac L r} }\lf_1,\;  \omega_{02}=-\frac{\dot \epsilon r}{2\sqrt{\epsilon +\frac L r} } \lf_2, \; \omega_{03} = \frac{\dot \epsilon L}{2(\epsilon +\frac L r)^{\frac 3 2  }}\lf_3, \; \omega_{04} = -\frac{\dot \epsilon}{2\sqrt{\epsilon +\frac L r} }dr. \nonumber \\
\end{align}

In terms of the left-invariant vector fields     $X_1,X_2,X_3$ on $SU(2)$  which are dual to the one-forms $\sigma_1,\sigma_2, \sigma_3$, 
the frame field dual to the co-frame \eqref{tetrad} is 
\bee
\label{vfields}
E_0=\frac{\partial}{\partial t},\quad 
E_1=\frac 1 a X_1, \quad  E_2=\frac 1 a  X_2\quad  E_3=\frac 1c  X_3, \quad E_4=\frac r  a \frac{\partial}{\partial r}.
\eee
 Then, with $A$ standing for the $t$-dependent one-form \eqref{connection} and 
 \bee
A_\mu =A(E_\mu), \qquad  \omega^{\kappa\lambda}_{\mu}=  \omega^{\kappa\lambda}(E_\mu),
\eee
the Dirac operator  is 
\bee
\label{gendirac}
\Dslash_{p,t} =  \gamma^{\mu}\left(E_{\mu} +A_\mu- \frac{1} {8} [\gamma_{\kappa},\gamma_{\lambda}] \omega^{\kappa\lambda}_{\mu}\right).
\eee
Inserting \eqref{GH} for $a$ and $c$ and the spin connection, this simplifies to 
\bee
\Dslash_{p,t}= 
\gamma^0\left(\frac{\partial}{\partial t}  + \frac{1}{2}\frac{\dot\epsilon r }{\epsilon r +L} \right) + \Dslash_p,
\eee
where $\Dslash_p$ is the four-dimensional Dirac operator \eqref{TNdirac} for  the metric $g_{\text TN}(t)$,  coupled to the connection \eqref{tconnection} at some given $t$. 

Rather remarkably, we can obtain  solutions of the time-dependent Dirac equation
\bee
\label{timeDirac}
\Dslash_{p,t}\Psi=0
\eee
 by taking zero-modes  of the form \eqref{holosol} with  adiabatically varying radial dependence 
\bee
\label{radsolt}
R_j(r,t )=C\frac{r^j}{ \sqrt{\epsilon(t) r +L} }e^{((2j+1)-p(t))\frac{\epsilon(t)  r}{2L} },
\eee
provided that 
 \bee
 \frac{\partial}{\partial t}  R_j(r,t )= - \frac{1}{2}\frac{\dot\epsilon r }{\epsilon r +L} R_j(r,t ).
 \eee
 This  holds provided $p$ is constant and satisfies
 \bee
 p=2j+1,
 \eee
 so that the exponential  factor in \eqref{radsolt} vanishes. 
 
 Thus,  we recover the quantisation condition on $p$ which ensures that   for  fixed $t$ the connection \eqref{tconnection}  extends to $\CP^2$ as the condition for zero-energy solutions of the time-dependent Dirac equation \eqref{timeDirac}. The resulting solutions have the radial dependence 
 \bee
 \label{tposq}
 R_j(r,t )=C\frac{r^j}{ \sqrt{\epsilon(t) r +L} },
 \eee
 and are not square-integrable over TN for any fixed  value of $t$. 
 
 Again,  the $j=1/2$ solution is special in that it tends to a constant at spatial infinity and is proportional to the fibre radius $2c$  for any value of $t$. 
 Extending the earlier definitions to hold for any fixed $t$
 \bee
 c(t,r)= \frac{L}{\sqrt{\epsilon(t) +\frac Lr }}, \qquad w(t,r,z) =c(t,r)z,
 \eee
 the formula \eqref{holosolw} for the spin 1/2 solution in terms of $w$ also  holds  in the time-dependent case: 
 \bee
\label{holosolt}
\Psi_{\frac 12 } (t,r,z_1,z_2) = \begin{pmatrix}
a_{-\frac 1 2} w_1 +  a_{\frac 1 2} w_2\\
0\\
0 \\
0
\end{pmatrix}.
\eee

The energy-momentum tensor for the spinor \eqref{holosolt} with radial and time-dependence \eqref{tposq} does not vanish. The metric \eqref{timeTN}, the gauge potential \eqref{tconnection} and the spinor \eqref{holosolt} with $(r,t)$-dependence  \eqref{tposq}  satisfy, respectively, the Einstein equations (provided $\epsilon$ is affine in $t$), the  Maxwell equations (provided, in addition, that $p$ is affine in $t$) and the Dirac equation (provided, in addition, that $p$ is constant and an integer); they do not satisfy the coupled Einstein-Maxwell-Dirac system.

\section{Geometric models of particles and Dirac's Large Number Hypothesis}
\label{concl}

The idea of describing particles in terms of everywhere smooth solutions of the Einstein equations was  explored  by  Einstein and Rosen almost 80 years ago  in \cite{ER}.  Implementing this idea in a realistic model of elementary particles is beset with various well-known difficulties, the most basic of which is the problem of scales. 

The problem is   illustrated by a close cousin of the model discussed in in this paper, namely the trivial (4+1)-dimensional extension of the static Taub-NUT geometry into a solution of the Einstein equations. In \cite{Sorkin,GP},  this spacetime is interpreted as a static  Kaluza-Klein monopole, and it is shown  that  the mass of the monopole is proportional to the ratio of the asymptotic fibre radius  to Newton's constant in 3+1 dimensions. As a result, the mass associated to the typical  length scale  $10^{-15}$ m of nuclear physics is a huge $10^{12}$ kg,  equivalent to   the mass of $10^{39}$  protons.

The time-dependent geometric model of a particle discussed  in this paper continues  the line of thought started by Einstein and Rosen but  needs to be interpreted with care in order   to avoid unrealistic relations between size and mass of the kind described above.  As in the discussion of static models in \cite{AMS},  we do this in the first instance by avoiding the problem of unification and aiming at a geometrisation of particles and their  non-gravitational interactions only. Also note that the models in \cite{AMS} and the time-dependent extension introduced here do not yet include a theory of  energy and dynamics, though proposals for both have been  put forward \cite{FM}. 

The TN model of the electron as proposed in   \cite{AMS}  only  captures basic properties of the electron: the negative of the electron charge $e$  through  the Chern class of the asymptotic $U(1)$ fibration, the point-like nature of the electron through the single fixed-point of the $U(1)$ action (the NUT),  and the mass  of the electron trough the interpretation of  the asymptotic length  $\Lambda$ of the $U(1)$ fibre as the classical electron radius. The electron mass, in units where the speed of light is one, is  therefore
\bee
\label{emass}
m_e=\frac{e^2}{\Lambda}.
\eee

An additional insight gained in the current paper  is that TN space is naturally interpreted as the  spin space of the electron in the sense that spin 1/2 zero-modes of the twisted Dirac operator are linear functions of the microscopic complex coordinate \eqref{wcoord}  defined in terms of the $U(1)$  action and its orbit length. Thus the beautiful link between spin 1/2 and the unit charge  Dirac monopole via the Hopf fibration finds a natural place in the TN model of the electron. 

Having arrived at this interpretation through the study of conventional spinors obeying the Dirac equation on TN, it is tempting to discard the Dirac operator and to simply postulate that spin 1/2 states  of the electron are  the linear functions  of the microscopic complex TN coordinates. Either way, the resulting picture is one where the TN geometry  has two facets,  one  microscopic  and referring to electron's spin degrees of freedom, and the other macroscopic and referring to its position in Euclidean three-space. 

Put this way, the picture, already incorporating Dirac's monopole and equation, becomes an illustration of  yet another of Dirac's ideas, namely his  proposal of two metrics, made in the context of his Large Number Hypothesis (LNH).  

Briefly, the LNH states that large numbers in physics such as the ratio of the electromagnetic to gravitational force between proton and electron are of the same order of magnitude as  the age of the universe  in atomic units \cite{Dirac-1,Dirac0,Dirac1,Dirac2}. This necessarily requires one or more  of the fundamental constants of nature to vary with the age of the universe. Dirac proposed that Newton's constant  $G$ should vary while  the masses of elementary particles should remain constant. 

Since Einstein's theory of gravity requires a constant value for $G$, Dirac then postulated two systems of units, Einstein units in which Einstein's equations hold,  and atomic units in which $G$ varies with time and  Einstein's equations do not necessarily  hold. There are two distinct metrics, one for each set of units. 

It turns out that our microscopic and macroscopic  interpretation of the TN geometry   are illustrations of  Dirac's LNH and two-metric formalism. In fact, several aspects of the TN model may be viewed as precise versions of ideas proposed by Dirac in the context of the LNH, as we shall now explain.

Far from the  NUT, i.e., for  large $r$, the $U(1)$ fibre is negligible and the metric is essentially the flat metric $d\vec{x}^2$ of macroscopic  Euclidean space. Near the NUT, i.e.,  for small $r$,  the metric is the flat metric \eqref{atomic} on microscopic spin space. 
The relation between the macroscopic line element $dr$ and the microscopic line element $dc$ for small $r$ is, according to \eqref{rcrel}, 
\bee
Ldr \approx c dc.
\eee
This relation is nothing but the  spatial version of Dirac's formula, derived  using dimensional analysis
in   \cite{Dirac2},
 \bee
d\tau= tdt,
\eee
which relates the  time $\tau$  measured in Einstein units and the  time $t$ measured in atomic units. It is this similarity   which inspired the current paper. 
 
Dirac  argued that the masses of particles should be the same  whether they are expressed in atomic or Einstein units. In our  model, this is captured by the fact that the length scale $\Lambda$ appearing in the mass formula \eqref{emass}  is defined in terms of the asymptotic geometry but  
 can also be extracted from the geometry  at the NUT according to  \eqref{Gauss}. This is a non-trivial feature of the TN geometry which implies the equality  of `local' and `asymptotic' mass.

Finally, the time-dependent solution \eqref{timeTN} with  affine evolution of  $\epsilon$  captures  yet further aspects  of the LNH.  As we  saw, the metric satisfies the (4+1)-dimensional Einstein equations  and the abelian gauge  potential and spinor  satisfy the linear Maxwell and Dirac equations, but they do not satisfy the non-linear and fully coupled Einstein-Maxwell-Dirac equations. From Dirac's point of view, this is not surprising since the coupling between matter and geometry in atomic units needs not be of the usual Einstein form. 

The details of the  time evolution of length scales   in the time-dependent geometry \eqref{timeTN}  also   fit well with the LNH.
Recall from  \eqref{epdef}  that  
$\sqrt{\epsilon}$ is the ratio of two length scales, namely the macroscopic length scale $L$ for the base of TN and the microscopic length scale $\Lambda$ of the fibre.
Concentrating again on the case $\epsilon(t)=t$, the  resulting metric
\bee
ds^2=\left(t  + \frac {L}{r}\right) d\vec{x}^2 + \frac {L^2r}{tr+ L}\lf_3^2
\eee
has the asymptotic (large $r$)  form
\bee
ds^2  \approx  td\vec{x}^2 + \frac {L^2}{t} \lf_3^2.
\eee

To  compare  length scales  in the base and the fibre, we introduce
 dimensionless spatial coordinates $\vec{\xi}= \vec{x}/L$, matching the dimensionless angular coordinate $\gamma \in [0, 4\pi)$ on the fibre.  The metric  then reads
\bee
ds^2  \approx  t L^2d\vec{\xi}^2 + \frac {L^2}{t} \lf_3^2. 
\eee 
The macroscopic length scale $L_M$ for the base and the  microscopic length scale$L_m$ for the fibre are thus
\bee
L_M= \sqrt{t} L, \qquad L_m = \frac{L}{\sqrt{t}},
\eee
and so
\bee
\frac{L_M}{L_m} = t.
\eee

Identifying the order of magnitude of $L_M$ with the length scale associated to the current estimate  $10^{-52} {\text m}^{-2}$ for the cosmological constant,
\bee
L_M  = 10^{26} \,  \text{m},
\eee
and the order of magnitude of   $L_m$  with the classical electron radius
\bee
L_m= \frac{e^2}{m_e c^2} \approx 10^{-15} \, \text{m},
\eee
we indeed find that $ L_M/ L_m \approx 10^{41} $
is close to Dirac's large number.  
This suggests the interpretation of $t$ as the age of the universe in atomic units and thus the following tentative interpretation of the model as a geometric description of the electron's time evolution since the big bang.   

Identifying   $t=0$ with the time of  the big bang, the initial geometry is $\RR^4$, but for $t>0$ one dimension compactifies and becomes the  fibre in a dual Kaluza-Klein model of electromagnetism. The size of the internal dimension is initially  comparable to macroscopic scales, but it shrinks as time evolves. From this point of view, the reason why the  classical electron radius is  small relative to cosmological length scales today is simply that this shrinking  has  gone on for a long time. Note also that this model of a single electron comes with a preferred direction of time. As discussed after \eqref{arrow},  backward evolution  from $t=0$ leads to a singular spacetime while forward evolution is smooth.

To sum up, we developed the description of spin in a geometric model of a particle and proposed a precise way of including time. We showed that the  description of spin can naturally be extended to the time-dependent model and that the resulting picture captures several key aspects of Dirac's LNH in a mathematically precise way. 

A model of a single particle necessarily only has limited scope for phenomenology and it is therefore important to extend the ideas of this paper to multi-particle models. Multi-TN space is a natural arena for modelling  multi-electron systems and it was shown in \cite{GLP} that the time evolution discussed here  for TN can be extended to the multi-TN metric. The interacting dynamics of electrons and their spins could probably be discussed by exploiting and combining the index results in   \cite{Pope1}  with the proposals for interaction energies in \cite{FM}, but the details  have yet to be  worked out.

One would also like to extend both the description of spin and the inclusion of time to models of baryons. In \cite{AMS}, the Atiyah-Hitchin (AH) metric was proposed as a (static) model for the proton.   The AH model is rotationally symmetric but   only has an additional   $U(1)$ isometry  asymptotically.  It captures
 the extended nature of the proton through a `core' region where the $U(1)$ action is not defined.

In  light of the  essential role played by the $U(1)$ fibration in this  paper, it may be more natural to consider only Ricci-flat four-manifolds  having  a globally defined $U(1)$ action as models for static particles.  The   Gibbons-Hawking classification of the possible  fixed point sets as  either  isolated  `nuts' or  two-dimensional `bolts'  \cite{GH} suggests  that the former could describe electrons and  the latter   baryons. The  Euclidean Schwarzschild  metric and the Taub-bolt (or Page) metric  \cite{Page} would then be natural candidates for the neutron and the proton.  It is therefore interesting to find time-dependent versions of these models, ideally also including a description of spin, and to interpret them along the lines of this paper.

\vspace{0.5cm}

\noindent {\bf Acknowledgements} \;We thank Jos\'e Figueroa O'Farrill for  numerous discussions  and acknowledge support from the EPSRC  through a research grant. BJS also thanks Maciej Dunajski for discussions.

\end{document}